\documentclass[a4paper]{article}
\usepackage{indentfirst}
\usepackage{hyperref}
\usepackage{multirow}
\usepackage{latexsym,bm,amsmath,amssymb}
\usepackage{graphicx}
\usepackage{epsfig}
\usepackage{subfigure}
\usepackage{cite}
\usepackage{color}
\usepackage[top=1in, bottom=1in, left=1.25in, right=1.25in]{geometry}
\usepackage{slashed}
\usepackage{fancyhdr}
\usepackage{slashed}
\usepackage{makecell,rotating,multirow,diagbox}
\usepackage{tikz}
\usepackage{float}
\usepackage{appendix}
\usepackage{setspace}
\usepackage{geometry}
\usepackage{graphics}
\usepackage{multirow}
\usetikzlibrary{shapes.geometric, arrows}
\usetikzlibrary{trees}
\usetikzlibrary{decorations.pathmorphing}
\usetikzlibrary{decorations.markings}
\tikzset{
        photon/.style={decorate, decoration={snake}, draw=red},
        nucleon/.style={draw=black, postaction={decorate},
           decoration={markings,mark=at position .55 with{\arrow[draw=black]{>}}}},
        pion/.style={draw=blue, postaction={decorate},
        decoration={markings,mark=at position .55 with{\arrow[draw=blue]{}}}},
        sigma/.style={draw=black, postaction={decorate},
        decoration={markings,mark=at position .55 with{\arrow[draw=black]{}}}},
        link/.style    = { draw=black, double = white, line width = 1.8pt, double distance = 0.8pt , postaction={decorate},decoration={markings,mark=at position .55 with{\arrow[draw=black]{>}}}},
    }

\newcommand{\be}{\begin{equation}}
\newcommand{\ee}{\end{equation}}
\newcommand{\ba}{\begin{array}{c}} \newcommand{\ea}{\end{array}}
\newcommand{\bqa}{\begin{eqnarray}}
\newcommand{\eqa}{\end{eqnarray}}

\def\bea{\arraycolsep .1em \begin{eqnarray}}
\def\eea{\end{eqnarray}}

\def\s0#1#2{\mbox{\small{$ \frac{#1}{#2} $}}}
\def\0#1#2{\frac{#1}{#2}}

\begin{document}
\setcounter{topnumber}{10}
\setcounter{totalnumber}{50}
\title{On Lowest Lying $\frac{1}{2}^-$ Octet Baryons }
\maketitle
\begin{center}
{\sc
Chang Chen,$^{\dagger\,}$ \,
Wen-Qi~Niu,$^{\dagger\,}$ \,
Han-Qing~Zheng$^{\heartsuit\,,\star\,,}$\footnote{Corresponding author.}}
\\
\vspace{0.5cm}
\noindent{\small{$^\dagger$ \it  Department of Physics and State Key
Laboratory of Nuclear Physics and Technology,
 Peking University, Beijing 100871, P.~R.~China}}\\
\noindent{\small{$^\heartsuit$ \it  College of Physics, Sichuan University, Chengdu, Sichuan 610065, P.~R.~China}}\\
\noindent{\small{$^\star$ \it   Collaborative Innovation Center of
Quantum Matter, Beijing, Peoples Republic of China}}
\end{center}
\begin{abstract}
The recently proposed $N^*(890)$ $1/2^-$ baryon is studied in a flavor $SU(3)$ scheme with $K$ matrix unitarization, by fitting to  low energy cross section and phase shift data. It is found that $N^*(890)$ co-exists with low lying poles in other channels, which have been extensively discussed in the literature, though they  belong to different  octets, in the $SU(3)$ limit. Hence the existence of $N^*(890)$ is further verified. 
\end{abstract}

In recent studies, a  novel $1/2^-$ negative parity nucleon state named as $N^*(890)$ is suggested to exist, in the $S_{11}$ channel  $\pi$N scattering amplitude.~\cite{Wang:2017agd,Wang:2018gul,Wang:2018nwi} The pole is  
found by  using  the production representation~\cite{Zheng:2003rw,Zhou:2006wm,Zhou:2004ms,Xiao:2000kx,He:2002ut}
(see also Ref.~\cite{Yao:2020bxx} for a review). Later the existence of $N^*(890)$ is also confirmed by a $K$--matrix analysis,\cite{Ma:2020sym} and the $N/D$ studies.~\cite{Li:2021tnt,Li:2021oou} Since the existence of $N^*(890)$ does not seem to be well accepted yet in the physics community, further evidences need to be gathered, which may be done by making the study in  
flavor $SU(3)$ basis, by associating $N^*(890)$  with (sub-threshold) resonances investigated in other channels -- which is the main purpose of this note. 

There have been actually many such studies in the literature, and these studies mainly concentrated on the strangeness $s=-1$ sector.
For example, In Ref.~\cite{Oller:2000fj}, a unitarized  $SU(3)$ $O(p^1)$ $\chi$PT amplitude is used to study the meson--baryon scatterings in the $s=-1$ sector. An extension to the $O(p^2)$ level is performed in Ref.~\cite{Guo:2012vv} and the twin pole structure around $\Lambda^*(1405)$ (namely, $\Lambda^*(1405)$ and $\Lambda^*(1380)$) is found. At higher energies, a pole $\Lambda^*(1670)$ is also found. {The I=1 sector is also investigated} in Ref.~\cite{Guo:2012vv} and the $\Sigma^*(1620)$ and $\Sigma^*(1750)$ poles are found. Similar discussions are also made in Refs.~\cite{Oset:2001cn,Jido:2002yz,Jido:2003cb,Oller:2006jw}. In Ref.~\cite{Oset:2001cn} no negative parity $\Sigma$ resonances are found.
In Ref.~\cite{Jido:2003cb} two $\Sigma^*$ poles located at $1401-40i$ and $1488-114i$ are found,  the former mainly couples to $\pi\Sigma, \bar KN$ while the latter mainly couples to $\pi\Sigma$ and $K\Xi$. In Ref.~\cite{Oller:2006jw} two narrow $\Sigma$ poles at $1425-6.5i$ and $1468-13i$ are found with strong couplings to $\pi\Sigma$ channel.  
Furthermore, in Refs.~\cite{Gao:2010hy,Wu:2009tu}, it is suggested that, introducing an explicit $\frac{1}{2}^-$ $\Sigma(1380)$ resonance field in the effective lagrangian gives a better description to the {experimental data }
of $\gamma p\to K^+\Sigma^{0*}(1385)$, $\gamma n\to K^+\Sigma^{-*}(1385)$
and $K^-p\to \Lambda\pi^+\pi^-$.

Discussions on the possible existence of low lying $1/2^-$ $\Xi$ state can also be found in the literature. A Bethe--Salpeter (BS) equation approach is made in studying the $s=-2$ sector in Ref.~\cite{Garcia-Recio:2003ejq}. The $O(p^1)$ contact term extracted from chiral meson baryon interaction lagrangian is used as the BS kernel. Two $1/2^-$ $\Xi$ poles are found: $\Xi^*(1620)$ and $\Xi^*(1670)$, in addition to the twin pole structure near $\Lambda^*(1405)$. Finally a review on related physics may also be found in Refs.~\cite{Mai:2020ltx}, \cite{Ikeda:2012au}. See also Ref.~\cite{Azimov:2003bb} for related discussions.

In this letter, we try to assimilate these results on low lying $1/2^-$ resonances in different sectors into a unified picture through $SU(3)$ argument.  
We start by considering these (extra) $1/2^-$ resonances as dynamically generated ones, i.e., they do not appear as explicit degrees of freedom in the effective lagrangian which is to be unitarized and fit to the data. This picture
is different from the strategy adopted in, for example, Refs.~\cite{Gao:2010hy,Wu:2009tu}. We see that this picture gives a reasonable description to experimental data. Next we also discuss the possibility that these low lying resonances are explicit degrees of freedom written in the effective lagrangian. We find, however, that these poles put in by hand in the effective lagrangian flee away in the complex plane and play a little role in improving chi-square. Hence we conclude that these low lying resonances are dynamically generated ones, and are of molecular type.

The  $SU(3)$ $\chi$PT lagrangian describing the lowest lying $1/2^+$ baryon--meson interactions at $O(p^2$) level
are
\begin{equation}\label{1}
	\begin{aligned}
		{\cal L}^{(1)} &= \mathrm{Tr}[\bar{B}i\gamma^\mu[D_\mu,B]] -M_0\cdot \mathrm{Tr}[\bar{B}B] \\
		&+ D\cdot \mathrm{Tr}[\bar{B}\gamma^\mu\gamma^5\{u_\mu,B\}]
		+ F\cdot \mathrm{Tr}[\bar{B}\gamma^\mu\gamma^5[u_\mu,B]]\ ,
	\end{aligned}        
\end{equation}
\begin{equation}\label{2}
	\begin{aligned}
		{\cal L}^{(2)} = &b_0\langle \bar{B}B\rangle\langle\chi_+\rangle + b_D\langle\bar{B}\{\chi_+,B\}\rangle + b_F\langle\bar{B}[\chi_+,B]\rangle\\
		&+b_1\langle\bar{B}[u_\mu,[u^\mu,B]]\rangle+b_2\langle\bar{B}\{u_\mu,\{u^\mu,B\}\}\rangle\\
		&+b_3\langle\bar{B}\{u_\mu,[u^\mu,B]\}\rangle+b_4\langle\bar{B}B\rangle\langle u_\mu u^\mu\rangle + \cdots \ .
	\end{aligned}        
\end{equation}
The notations and symbols are standard, following for example, Refs.~\cite{Oller:2006yh},\cite{Chen:2012nx}.
The relations between baryon masses and $SU(3)$ parameters are listed in the following: 
\begin{equation}
	\begin{aligned}
		M_N &= M_0 - 2(b_0+2b_F)m_\pi^2 - 4(b_0+b_D-b_F)m_K^2\ ,\\
		M_\Lambda &= M_0 -2(b_0 -\frac{2}{3}b_D)m_\pi^2 - 4(b_0 +\frac{4}{3}b_D)m_K^2\ ,\\
		M_\Sigma &= M_0 - 2(b_0 +2b_D)m_\pi^2 - 4b_0m_K^2\ ,\\
		M_\Xi &= M_0 - 2(b_0-2b_F) m_\pi^2 -4(b_0+b_D+b_F)m_K^2\ ,
	\end{aligned}
\end{equation} where $m_\pi^2 = 2B_0 m_u$, $m_K^2 = B_0(m_u + m_s)$, and $m_\eta^2 = \frac{2}{3} B_0(m_u+2m_s)$.
 {Only three out of four relations given above are independent and can be used to fix three parameters on the $r.h.s.$, leaving only one ($M_0$) for free.}
 
\begin{figure}[h]
	\centering                 
	\subfigure{}                    
	\begin{minipage}{4cm}\centering                      
		\includegraphics[scale=0.35]{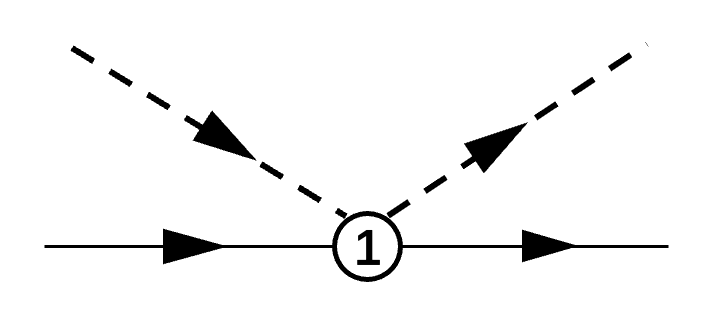}             
	\end{minipage}
	\subfigure{}                    
	\begin{minipage}{4cm}
		\centering                                   
		\includegraphics[scale=0.35]{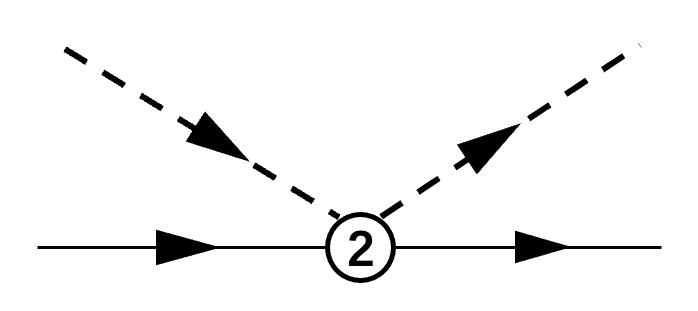}           
	\end{minipage}
	\caption{Contact $O(p^1)$ and  $O(p^2)$ diagrams.}\label{diag1}                                         
\end{figure}

\begin{figure}[h]
	\centering                 
	\subfigure{}                    
	\begin{minipage}{4cm}\centering                      
		\includegraphics[scale=0.4]{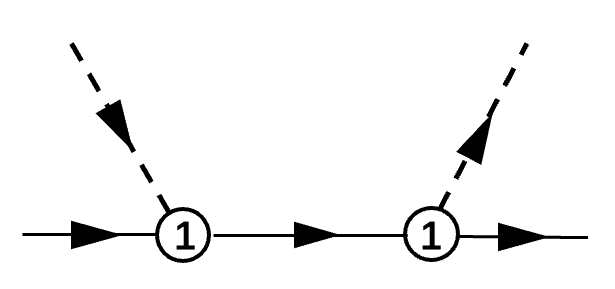}             
	\end{minipage}
	\subfigure{}                    
	\begin{minipage}{4cm}
		\centering                                   
		\includegraphics[scale=0.4]{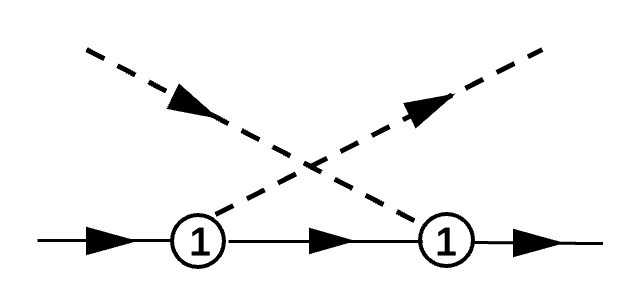}           
	\end{minipage}
	\caption{s and u channel diagrams at $O(p^1)$ level.}\label{diag2}                                         
\end{figure}

 The Feynman diagrams describing meson baryon $\to$ meson baryon scatterings at $O(p^2)$ level, generated by lagrangian Eqs.~(\ref{1}) and (\ref{2})  are depicted in Figs.~\ref{diag1} and \ref{diag2}.
We use a simple $K$-matrix unitarization scheme to improve the $O(p^2)$ perturbation amplitudes. A useful auxiliary function $\mathbf{g}(s)\equiv \mathrm{diag}\{g_i(s)\}$ is introduced, 
\begin{equation}\label{gis}
	g_i(s) = \int \frac{d^4q}{(2\pi)^4} \frac{1}{(q^2-M_i^2+i\epsilon)((P-q)^2-m_i^2+i\epsilon)},\,\, (s = P^2)
\end{equation}
where 
$M_i$ is the baryon mass and $m_i$ the meson mass in the $i$-th channel.
The expression of $g_i(s)$ in Eq.~(\ref{gis}) is renormalized using  standard dimensional regularization, which introduces an explicit renormalization scale ($\mu$) dependence (see for example, \cite{Gasser:1987rb}). Notice that in our fit we choose different $\mu$ parameters in  channels with different strangeness numbers. Though it is expected these $\mu$ parameters do not differ much.
{In practice we use the relation:
\begin{equation}
	 \mathbf{T}^{-1} = \mathbf{K}^{-1} -  \mathbf{g}(s),
\end{equation}
where $\mathbf{K}$ is the tree level $S_{11}$ channel scattering amplitude,} and $\mathbf{T}$
the unitarized scattering $T$ matrix. 

\begin{figure}[h]
	\centering
	\includegraphics[scale = 0.5 ]{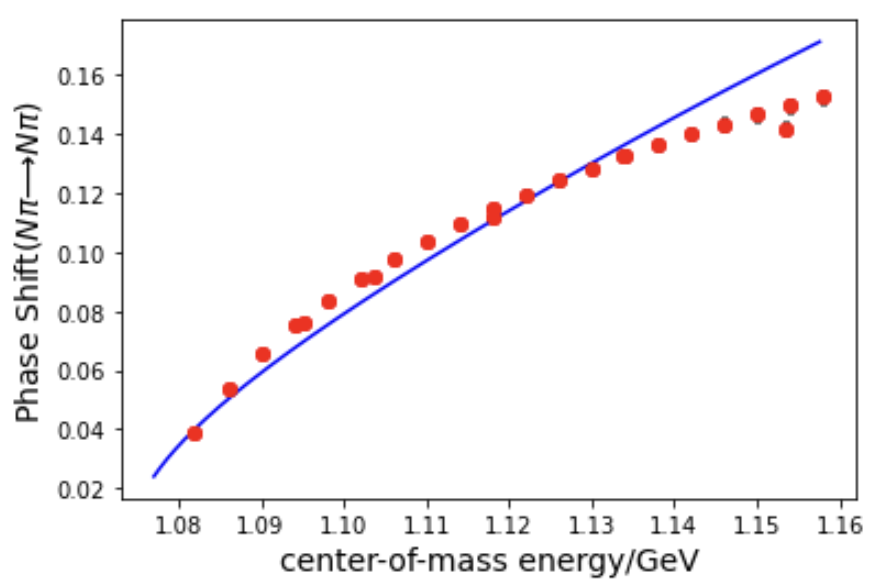}\\
	\caption{The fit to phase shift data in the region $\sqrt{s}= 1.08 - 1.16$GeV (data from Ref.~\cite{Hoferichter:2015hva}).}\label{fig1}
\end{figure}

\begin{figure}[h]
	\centering
	\includegraphics[scale = 0.45 ]{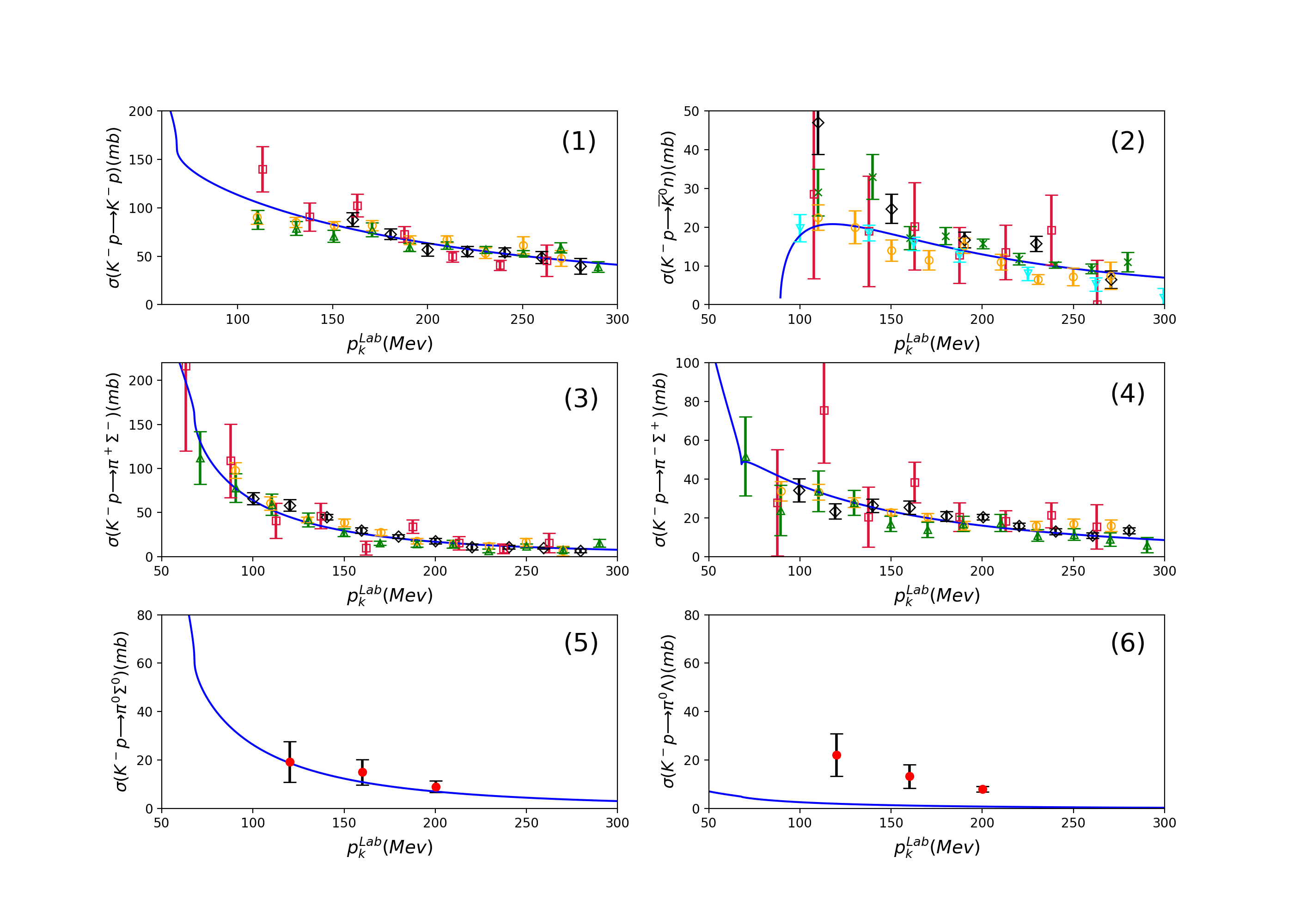}\\
	\caption{Fit to the experimental data in the $s = -1$ sector. The data points represented by black diamond, crimson square, orange circle, green cross, cyan down-triangle and green up-triangle in the first four panels are taken from Refs.~\cite{Ciborowski:1982et}\cite{Humphrey:1962zz}\cite{Kim:1967zze}\cite{Evans:1983hz}\cite{Kittel:1966zz}\cite{PhysRev.139.B719}, respectively. The data in the fifth and sixth panels are from Ref.~\cite{Martin:1980qe}.
	}\label{fig2}
\end{figure}

The processes under concern are $N\pi \to N\pi$, $K^-p \to K^-p$, $K^-p \to \overline{K}^0 n$, $K^-p \to \pi^+\Sigma^-$, $K^-p \to \pi^-\Sigma^+$, $K^-p \to \pi^0\Sigma^0$, $K^-p \to \pi^0\Lambda$.
{In the beginning we start with a $2\times2$ $I=1/2$ matrix amplitude fit for $\pi N$ and $N\eta$ channels, a $3\times 3$ $I=1$ matrix amplitude fit for $\Lambda\pi$, $\Sigma\pi$ and $N\bar K$ channels, and a $2\times2$ $I=0$ matrix amplitude fit for $\Sigma\pi$, $N\bar{K}$ channels, and neglecting higher thresholds.} The fit parameters are listed in Tab.~\ref{Tab1} and the fit curves are plotted in Figs.~\ref{fig1} and \ref{fig2}. In the fit the $s$--wave approximation is used.
\begin{table}[htbp]
	\centering
		\begin{tabular}{lc}
		\hline
		& $\chi^2_{d.o.f} = 5.1$  \\
		\cline{1-2}
		D & 0.4 \\
		F & 0.23 \\
		$f_0(\rm{MeV})$& 103.8 \\
		$M_0(\rm{GeV})$& 1.1  \\
		$b_0(\rm{GeV^{-1})}$& -0.044\\
		$b_D(\rm{GeV^{-1})}$&  0.026\\
		$b_F(\rm{GeV^{-1})}$& -0.189\\
		$b_1(\rm{GeV^{-1})}$& $0.647\pm 0.048$\\
		$b_2(\rm{GeV^{-1})}$& $0.672\pm 0.062$ \\
		$b_3(\rm{GeV^{-1})}$& $-0.121\pm 0.016$ \\
		$b_4(\rm{GeV^{-1})}$& $-0.701\pm 0.063$ \\
		$\mu_{S=0}(\rm{GeV)}$ &$0.703\pm 0.053$\\
		$\mu_{S=-1}(\rm{GeV)}$ &$0.727\pm 0.008$\\
		\hline
	\end{tabular}
\caption{Fit parameters from lagrangian Eqs.~(\ref{1}), (\ref{2}). Parameters without error bars are fixed ones.}\label{Tab1}
\end{table}
It is noticed that the main contribution to the chi-square comes from the fit to the $S_{11}$ channel phase shift data which contain very small error bars.\footnote{Here we follow a similar strategy to that of Ref.~\cite{Wang:2017agd} and define an error assigned to every point as the sum in quadrature of a systematic plus a statistical error, $err(\delta) = \sqrt{e_s^2+e_r^2\delta^2}$, where $e_s(=0.1^\circ)$ is the systematic error and $e_r(=2\%)$ the relative one.}


\begin{figure}[h]
 	\centering
 	\includegraphics[scale = 0.5 ]{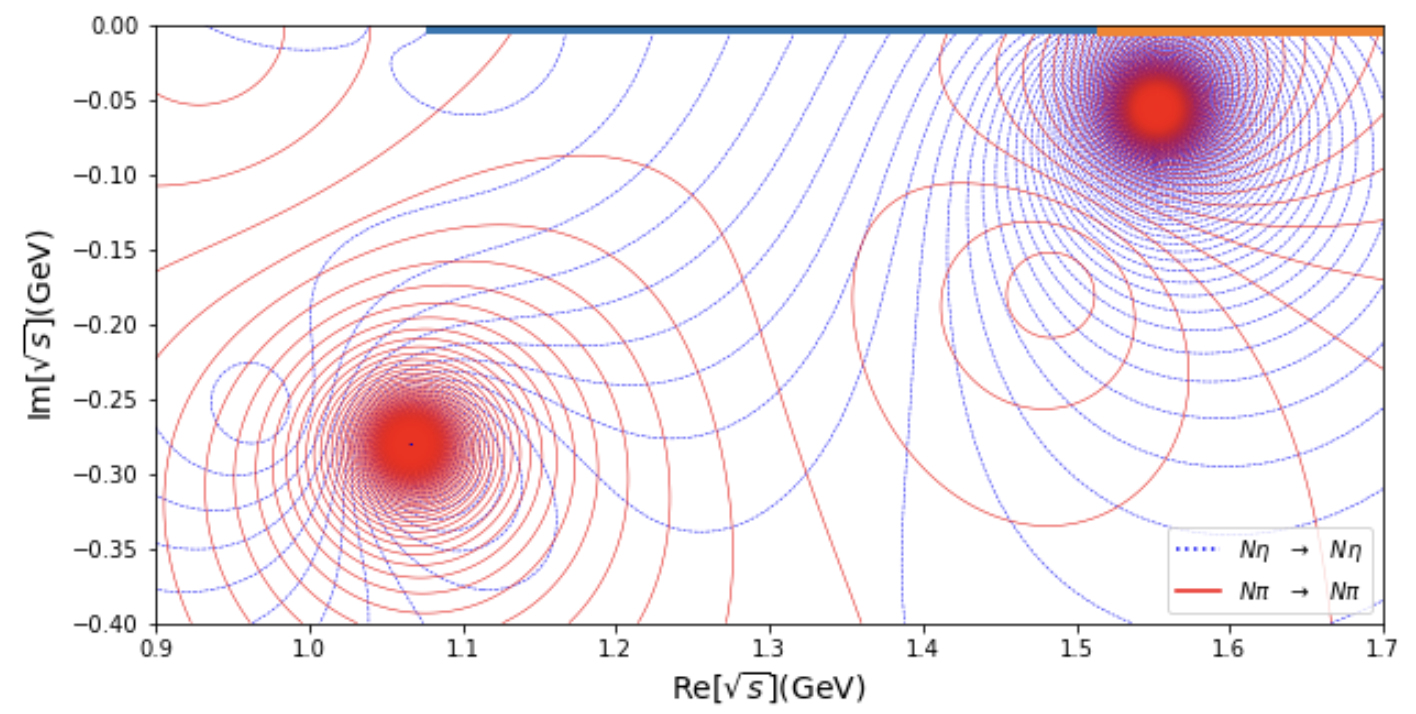}\\
 	\caption{The $N^*$(890) and $N^*$(1535) poles in sheet {(-,+).} The corresponding thresholds are marked with thick lines in the upper edge of the box.}\label{fig890}
 \end{figure}

\begin{figure}[h]
	\centering
	\includegraphics[scale = 0.5 ]{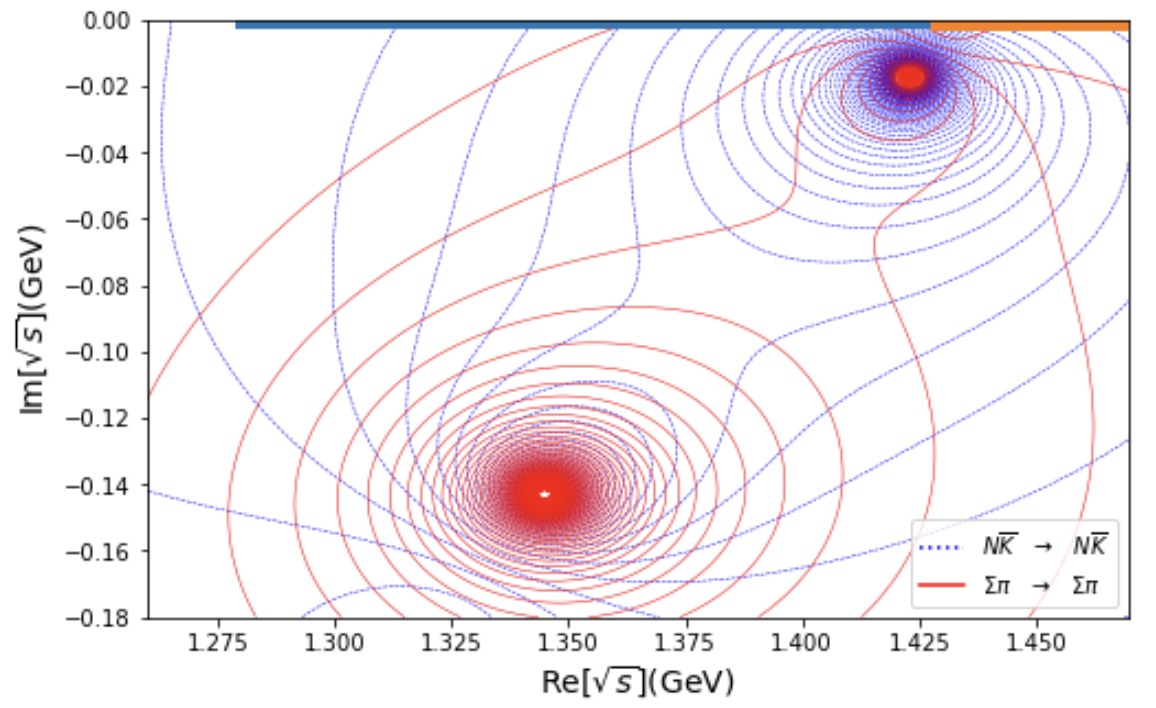}\\
	\caption{$\Lambda^*$(1405) and $\Lambda^*$(1380) in sheet (-,+).}\label{fig3}
\end{figure}

 In the amplitudes poles on different sheets are searched for.
 In the $N\pi$, $N\eta$ channel, it is clearly seen from Fig.~\ref{fig890} the pole location of the wanted $N^*(890)$ and $N^*(1535)$, and the precise location of two poles and their couplings are listed in Tab.~\ref{tab2}.
 \begin{table}[h]
	\centering  
\begin{tabular}{|c|c|c|c|}  
  \hline  
  Pole&  location&$|g_{N\pi}|$&$|g_{N\eta}|$\\
  \hline  
  $N^*$(890)& $1.066- 0.280 i $&0.617&0.436\\  
  \hline
  $N^*$(1535)&$1.553- 0.056 i$&0.645&1.031\\
  \hline
\end{tabular}
\caption{Pole locations and channel couplings in the $s = 0$, I = ${1}/{2}$ sector. All numbers in the table (and hereafter) are in units of GeV.} \label{tab2}
\end{table}
Meanwhile, results in the $s=-1, I=0$ ($N\bar{K}$, $\Lambda\eta$) sector are listed in Fig.~\ref{fig3} and Tab.~\ref{tab3}. 
\begin{table}[h]
	\centering  
	\caption{Pole locations and channel couplings in the $s = -1, I = 0$ sector.}\label{tab3}  
	\begin{tabular}{|c|c|c|c|}  
		\hline  
		Pole&  location&$|g_{\Sigma \pi}|$&$|g_{N\overline{K}}|$\\
		\hline  
		$\Lambda^*$(1380)&$1.345- 0.143 i$&1.032&0.702 \\ 
		\hline
		$\Lambda^*$(1405)&$1.423- 0.017 i$&0.453&0.966 \\
		\hline
	\end{tabular}
\end{table} 
It is worth emphasizing that in the present $2\times2$ fit, there are only two poles as shown in Tab.~\ref{tab3}. This ``twin pole" structure have been
previously extensively discussed in the literature and our results may be considered as an further confirmation. Moreover, when one goes to the complete $4\times 4$ fit by including $\Lambda\eta$ and $\Xi K$ channels, another $1/2^-$ $\Lambda$ baryon resonance appears at $\sqrt{s}=1.81-i0.04$GeV, {in qualitative agreement with the $\Lambda(1670)$ pole found in Ref.~\cite{Guo:2012vv}.}

\begin{figure}[h]
	\centering
	\includegraphics[scale = 0.45 ]{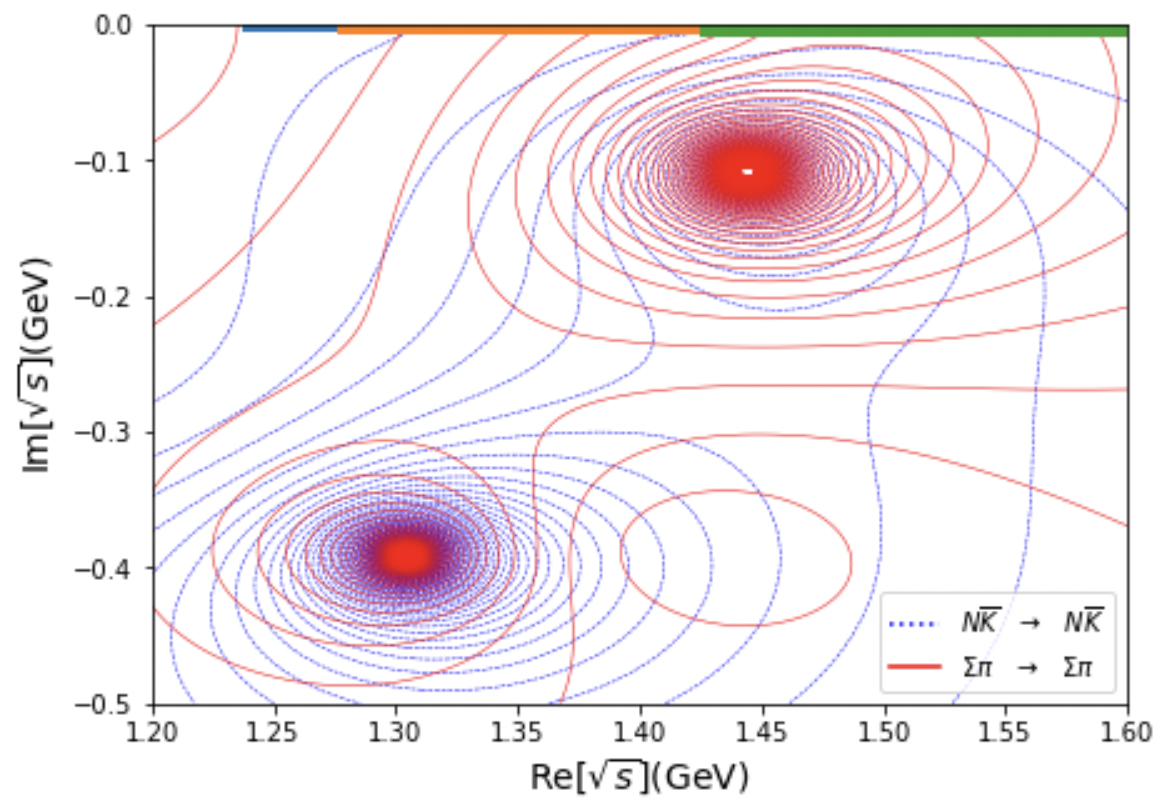}\\
	\caption{$\Sigma^*$(1360) and $\Sigma^*$(1620) poles  in sheet (-,-,+).}\label{fig5}
\end{figure}

Another $3\times3$ fit in the $s=-1, I=1$ sector ($\Sigma\pi$, $N\bar{K}$, $\Lambda\eta$) are depicted in Fig.~\ref{fig5} and Tab.~\ref{tab5}.
\begin{table}[h]
	\centering  
	\caption{Pole locations and channel couplings  in the $s = -1, I = 1$ sector.}\label{tab5}  
	\begin{tabular}{|c|c|c|c|}  
		\hline  
		Pole&  location&$|g_{\Sigma \pi}|$&$|g_{N\overline{K}}|$\\
		\hline  
		$\Sigma^*$(1380)& $1.305- 0.392 i$&0.578&1.346 \\ 
		\hline
		$\Sigma^*$(1650)& $1.444- 0.109 i$&0.940&1.084\\
		\hline
	\end{tabular}
\end{table}
It is remarkable to notice that the results listed in Tab.~\ref{tab5}
are in agreement with the result of Refs.~\cite{Jido:2003cb,Oller:2006jw}.

\begin{figure}[h]
	\centering
	\includegraphics[scale = 0.45 ]{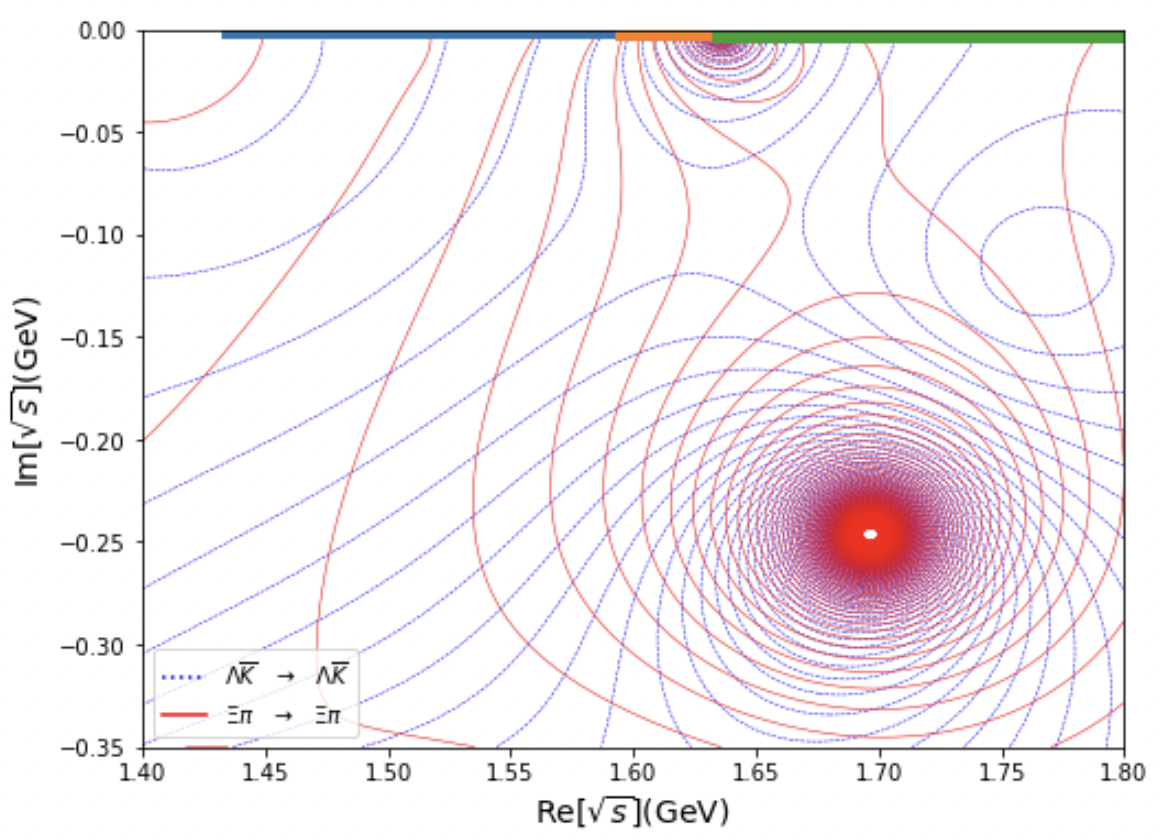}\\
	\caption{Pole locations and channel couplings in the $s = -2, I=1/2$ sector, in sheet (-,+,+).}\label{fig6}
\end{figure}

At last, we also listed in the Fig.~\ref{fig6} and Tab.~\ref{tab6} the results
in the $s=-2$ sector. Here we analyze a triple channel amplitude ($\Xi\pi,\Lambda\bar{K},\Sigma\bar{K}$), in accordance with Ref.~\cite{Ramos:2002xh}. \footnote{Notice that there is no fit in the $s=-2$, sector.
Fig.~\ref{fig6} and Tab.~\ref{tab6} are obtained by taking $\mu(s=-2)=2\mu(s=-1)-\mu(s=0)$.}  Notice that in Tab.~~\ref{tab6} one of the pole contains a too small width. But we think this is not worrisome since the choice of the $\mu$ parameter is somewhat arbitrary. The most important fact is that there are two $\Xi^*$ poles which coincides with the result of Ref.~\cite{Garcia-Recio:2003ejq}.
\begin{table}[htbp]
	\centering   
	\caption{ Pole locations and channel couplings in the in $s= -2, I = {1}/{2}$ sector.}\label{tab6}   
	\begin{tabular}{|c|c|c|c|}  
		\hline   
		Pole&  location&$|g_{\Xi \pi}|$&$|g_{\Lambda \overline{K}}|$\\
		\hline   
		Pole1& $1.636-0.0001i$&0.262&0.335\\  
		\hline
		Pole2& $1.696-0.246i$&1.087&1.208\\
		\hline
	\end{tabular}
\end{table}

It is seen from Fig.~\ref{fig2} that the last diagram (the fit to the $\sigma(K^-p\to\pi^0\Lambda)$) is not fit well. This can be improved in another fit solution, at the cost that the fit to the $S_{11}$ phase shift gets slightly worse, and that 
the $N^*(1535)$ pole and one of the $\Xi^*$ pole in Fig.~\ref{fig6} flee  away. Nevertheless, the existence of $N^*(890)$, as well as the ``twin pole structure" of $\Lambda^*$ and $\Sigma^*$ are not influenced.

 Notice that all the $J^P=1/2^-$ baryons discussed above are generated from an effective lagrangian without them. That is to say, these negative parity baryons are generated dynamically. In order to seriously address the question whether these states are of ``elementary" nature or only  hadronic molecules, one probably needs to make a study starting from an effective lagrangian with built-in negative parity baryonic fields. For this purpose, we write:
 \begin{equation}\label{Eq21}\begin{aligned}
 {\cal L} &=\operatorname{Tr}\left[\bar{B} i \gamma^{\mu}\left[D_{\mu}, B\right]\right]-M_{0} \cdot \operatorname{Tr}[\bar{B} B] \\
 	&+ \operatorname{Tr}\left[\bar{B}_- i \gamma^{\mu}\left[D_{\mu}, B_{-}\right]\right]-M^{*}_{0} \cdot \operatorname{Tr}[\bar{B}_- B_{-}] \\
 	&+D \operatorname{Tr}\left[\bar{B} \gamma^{\mu} \gamma^{5}\left\{u_{\mu}, B\right\}\right]+F \operatorname{Tr}\left[\bar{B} \gamma^{\mu} \gamma^{5}\left[u_{\mu}, B\right]\right]\\
 	&+D_{-} \operatorname{Tr}\left[\bar{B}_- \gamma^{\mu} \gamma^{5}\left\{u_{\mu}, B_{-}\right\}\right]+F_{-} \operatorname{Tr}\left[\bar{B}_- \gamma^{\mu} \gamma^{5}\left[u_{\mu}, B_{-}\right]\right]\\
 	& + D_{1}\left\{ \operatorname{Tr}\left[\bar{B}_- \gamma^{\mu} \left\{u_{\mu}, B\right\}\right]+ \operatorname{Tr}\left[\bar{B} \gamma^{\mu} \left\{u_{\mu}, B_{-}\right\}\right]\right\} \\
 	&+ F_{1}\left\{ \operatorname{Tr}\left[\bar{B}_- \gamma^{\mu} [u_{\mu}, B]\right]+ \operatorname{Tr}\left[\bar{B} \gamma^{\mu} [u_{\mu}, B_{-}]\right]\right\}\ ,\\
 \end{aligned}
\end{equation}
\begin{equation}\label{Eq22}
\begin{aligned}
	{\cal L}^{1}_{SB} =& b_{0}\langle\bar{B} B\rangle\left\langle\chi_{+}\right\rangle+b_{D}\left\langle\bar{B}\left\{\chi_{+}, B\right\}\right\rangle+b_{F}\left\langle\bar{B}\left[\chi_{+}, B\right]\right\rangle \\
	+& b^{'}_{0}\langle\bar{B}_- B_{-}\rangle\left\langle\chi_{+}\right\rangle+b^{'}_{D}\left\langle\bar{B}_-\left\{\chi_{+}, B_{-}\right\}\right\rangle+b^{'}_{F}\left\langle\bar{B}_-\left[\chi_{+}, B_{-}\right]\right\rangle\ ,\\
\end{aligned}
\end{equation}
\begin{equation}\label{Eq23}
\begin{aligned}
	{\cal L}^{2}_{SB} =  & e (\langle\bar{B}\chi_{-}B_{-}\rangle - \langle\bar{B}_-\chi_{-} B\rangle)\\
	&+h (\langle\bar{B} B_{-}\chi_{-}\rangle -  \langle\bar{B}_-B\chi_{-}\rangle)\ ,\\
\end{aligned}
\end{equation}
where $B_-$ denotes a $1/2^-$ baryon octet, and ${\cal L}_{tot.} ={\cal L} + {\cal L}^{1}_{SB} + {\cal L}^{2}_{SB}$. {In the fit, however, we find that including negative parity baryons explicitly in the effective lagrangian does not improve the fit quality much. Furthermore the  locations of the newly introduced poles are very unstable and have little influence to the locations of the dynamical poles as discussed in Figs.~\ref{fig890} --\ref{fig6}. Therefore we think that it is not necessary to include by hand extra ``elementary" negative baryon fields, in disagreement with some claims found in the literature.}

{At last, one needs to discuss the fate of those poles in the $SU(3)$ limit for the purpose of pinning down the property of the desired $N^*(890)$. To study this, it is realized that \cite{Bruns:2021krp} the $SU(3)$ symmetry
only exists in the sheet with same sign, i.e., in sheet $(+,+,\cdots,+)$
or sheet $(-,-,\cdots,-)$, and all channels should be taken in to considerations. Taking the $N^*(890)$ for example, there were only two channels being considered (i.e., $N \pi$, $N\eta$), now one has to  go to four channels ($N\pi$, $N\eta$, $\Lambda K$, $\Sigma K$) in sheet {$(-,-,-,-)$}. In our strategy, we first take the $SU(3)$ limit, $m_K^2\to m_\pi^2$ and trace the trajectories, starting from the sheet where the pole locates, i.e, sheet $(-,+,+,+)$. Then we multiply the channel phase space factors $\rho_i$ ($i=2,3,4$)  a factor $k$, and let $k$ changes smoothly from $+1$ to $-1$. In this way one can trace the trajectory of the $N^*(890)$ pole moving from the physical location to its destiny in the $SU(3)$ limit, in sheet (-,-,-,-).  
We find that  $N^*(890)$ and $N^*(1535)$ have different destinations, i.e., they belong to different octets. This is not surprising as it is known that in $SU(3)$ limit there exist actually two octets, as $8\otimes 8=1\oplus8\oplus8\oplus10\oplus\bar{10} \oplus27$. We do not trace other pole trajectories any more, as they are model and scheme dependent, and have been extensively discussed already in the literature.

In this note, we have carefully investigated the possible correlations between the newly proposed $N^*(890)$ resonance and those having been discussed extensively in the literature, e.g., $\Lambda^*(1405)$ and $\Lambda^*(1380)$.
Since they exist on the same footing, and the negative parity $\Lambda$
baryons are well accepted~\cite{ParticleDataGroup:2020ssz}, 
 there is little doubt, to the authors, on the existence of the $N^*(890)$.
\vspace{1cm}

$Acknowledgements:$ We would like to thank  Z.~H.~Guo for a careful reading of the manuscript and very helpful discussions.  This work is supported in part by National Nature
Science Foundations of China under Contract Number 11975028 and
10925522.
\renewcommand\refname{Reference}
\bibliographystyle{h-physrev}
\bibliography{NegativeB}
\end{document}